\documentclass[prd,aps,showpacs,superscriptaddress,eqsecnum,amsfonts,amsmath,twocolumn,preprintnumbers,floatfix,floats,nofootinbib,]{revtex4}
\usepackage{amsfonts,amsmath,units,wasysym,epsfig,graphicx,verbatim,color,
subfigure,epstopdf,bm}
\usepackage[colorlinks=true]{hyperref}

\definecolor{NTNUBlue}{rgb}{0.0470,0,0.5294}
\definecolor{Brown}{rgb}{0.5,0.5,0.5}
\definecolor{BrickRed}{cmyk}{0,.89,.94,.28}
\definecolor{FG}{cmyk}{0.75,0,1.0,0.5}

\begin{document}
%%%%%%%%%%%%%%%%%%%%%%%%% define some new commands %%%%%%%%%%%%%%%%%%%%%%%%%%%
\newcommand{\rhat}{\hat{r}}
\newcommand{\iotahat}{\hat{\iota}}
\newcommand{\phihat}{\hat{\phi}}
\newcommand{\h}{\mathfrak{h}}
\newcommand{\be}{\begin{equation}}
\newcommand{\ee}{\end{equation}}
\newcommand{\ber}{\begin{eqnarray}}
\newcommand{\eer}{\end{eqnarray}}
\newcommand{\UO}{Department of Physics,
1274 University of Oregon, Eugene, OR 97403-1274, U.S.A\\}
\newcommand{\WSU}{Department of Physics \& Astronomy, Washington State University,
1245 Webster, Pullman, WA 99164-2814, U.S.A \\}
\newcommand{\LIGOCaltech}{LIGO Laboratory, California Institute of Technology, 
Pasadena, CA 91125, U.S.A}
\newcommand{\TAPIR}{Theoretical Astrophysics, California Institute of Technology, 
Pasadena, CA 91125, U.S.A}
\newcommand{\IUCAA}{Inter-University Centre for Astronomy and Astrophysics, Post Bag 4, Ganeshkhind, Pune 411 007, India}

\title{Measuring neutron-star ellipticity with measurements of the stochastic gravitational-wave background} 

\author{Dipongkar Talukder}
\email{talukder@uoregon.edu}
\affiliation{\UO}

\author{Eric Thrane}
%\email{ethrane@ligo.caltech.edu}
\affiliation{\LIGOCaltech}

\author{Sukanta Bose}
%\email{sukanta@wsu.edu}
\affiliation{\WSU}
\affiliation{\IUCAA}

\author{Tania Regimbau}
\affiliation{UMR ARTEMIS, CNRS,
University of Nice Sophia-Antipolis, Observatoire de la C\^{o}te d'Azur, CS 34229 F-06304 NICE, France}

\date{\today}


\begin{abstract}
  Galactic neutron stars are a promising source of gravitational waves in the analysis band of detectors such as LIGO and Virgo.
  Previous searches for gravitational waves from neutron stars have focused on the detection of individual neutron stars, which are either nearby or highly non-spherical.
  Here we consider the stochastic gravitational-wave signal arising from the ensemble of Galactic neutron stars.
  Using a population synthesis model, we estimate the single-sigma sensitivity of current and planned gravitational-wave observatories to average neutron star ellipticity $\epsilon$ as a function of the number of in-band Galactic neutron stars $N_\text{tot}$.
  For the plausible case of $N_\text{tot}\approx 53000$, and assuming one year of observation time with colocated initial LIGO detectors, we find it to be $\sigma_\epsilon=2.1\times10^{-7}$, which is comparable to current bounds on some nearby neutron stars.
  (The current best $95\%$ upper limits are $\epsilon\lesssim7\times10^{-8}$.)
  It is unclear if Advanced LIGO can significantly improve on this sensitivity using spatially separated detectors.
  For the proposed Einstein Telescope, we estimate that $\sigma_\epsilon=5.6\times10^{-10}$.
  Finally, we show that stochastic measurements can be combined with measurements of individual neutron stars in order to estimate the number of in-band Galactic neutron stars. In this way, measurements of stochastic gravitational waves provide a complementary tool for studying Galactic neutron stars.
\end{abstract}

%\preprint{[LIGO-P1300134]}

\maketitle

\section{Introduction}
Of the estimated $10^{8}-10^{9}$ neutron stars in the Milky Way~\cite{refId0}, approximately $50,000$ are expected to rotate with ${\cal O}(\unit[]{ms})$ periods~\cite{1995ApJ...439..933L}.
Neutron stars with periods $T<\unit[200]{ms}$ emit gravitational waves (GWs)~\cite{PhysRevD.58.063001,PhysRevD.69.082004,Maggiore:1900zz,Regimbau:2001kx,Regimbau:2002ft} in the $\sim10$--$\unit[2000]{Hz}$ analysis band of current GW detectors such as LIGO~\cite{iligo} and Virgo~\cite{Virgoupdate}.
GW observatories have placed limits on GW emission from known pulsars~\cite{S5-Crab,S5-known,S5-Vela,S6-known}, from nearby neutron stars with unknown phase evolution~\cite{PhysRevLett.107.271102,S5-CasA}, and from electromagnetically quiet neutron stars \cite{S5-E@H1,S5-E@H2,S5-allsky1,S5allsky2}.
For nearby pulsars, direct GW searches have bounded neutron star ellipticities to be as low as $\epsilon\lesssim7\times10^{-8}$ at 95\% confidence level (CL)~\cite{S5-known}.
With the imminent arrival of second-generation GW detectors, the first detection of GWs from neutron stars might be just around the corner.
Even so, it is likely that the vast majority of Galactic neutron stars are too far away to observe individually in the near future.

Nonetheless, it may be possible to observe a stochastic signal~\cite{PhysRevD.59.102001} from the superposition of weak gravitational wave signals from the many Galactic neutron stars that are too far away to detect individually.
In this paper we show how measurements of the stochastic signal from Galactic neutron stars provide constraints that are independent and complementary to those derived from searches for individual neutron stars. Stochastic measurements of Galactic neutron stars provide more than just a cross-check for measurements of individual neutron stars---though, a robust model-independent cross-check is, in and of itself, useful.
By combining stochastic measurements with measurements of individual neutron stars, it is possible to gain insights into the ensemble properties of Galactic neutron stars, which are not otherwise accessible.
For example, one can estimate the total number of in-band neutron stars in the Milky Way.

The remainder of this paper is organized as follows.
In Sec.~\ref{sec:sourcemodel}, we describe models of Galactic neutron stars that can be employed by a stochastic search.
In Sec.~\ref{sec:methodology}, we discuss the methods used to estimate average neutron star ellipticity from stochastic signal measurements. Then, in Sec.~\ref{sec:searchresults}, we estimate the sensitivity of various GW observatories (both past and future) to stochastic signals from Galactic neutron stars.
We show how stochastic observations can be combined with observations of individual neutron stars to constrain the number of in-band neutron stars in the Milky Way.
In Sec.~\ref{sec:summary}, we conclude by summarizing prospects for future work.

In the appendix we discuss alternative analyses for deriving constraints on populations of neutron stars using the stochastic superposition of neutron star signals from (A) the Virgo Cluster and (B) the entire universe.
We argue that the stochastic signal from neutron stars in the Milky Way is stronger than either of these alternative sources, and therefore yields the most interesting constraints. Using the Virgo Cluster as a case study, we demonstrate how to measure neutron star ellipticity using the stochastic signal from an anisotropic source.
This methodology may be useful for future searches taking into account the anisotropic distribution of Milky Way neutron stars.

\section{Source model}\label{sec:sourcemodel}
In order to describe the stochastic signal from Galactic neutron stars, we model their distribution in both space and frequency.
We do not aspire to achieve a high degree of accuracy with our model, but only to sketch the qualitative features of the stochastic signal from Galactic neutron stars.
We revisit our assumptions in Section~\ref{sec:summary} and discuss how the results might vary for a more realistic model.

Our starting point is a population synthesis model. Following the formalism of~\cite{Story:2007xy}, we derive the distribution of neutron star period by evolving a set of simulated neutron stars with ${\cal O}(\unit[]{ms})$ periods from the end of the spin up phase to the present. We make the following assumptions: (i) uniform distribution of age $t$ between $0$--$\unit[12]{Gyr}$ (ii) log-uniform distribution of the initial magnetic field between $10^8$--$\unit[10^{12}]{G}$ (and no magnetic field decay) (iii) the initial period $P_0$ is assumed to match with the spin-up period derived from the formula
\begin{equation}\label{eq:ini_period}
P_{0} = 0.18 \times 10^{3\delta/7} B_{0}^{6/7}\,,
\end{equation}
where $P_{0}$ is in ms and $B_{0}$ is the initial magnetic field in units of $10^{8}$~G. The parameter $\delta$ is selected from a ramp distribution that increases by a factor $4$ between $0$--$2.8$~\cite{Story:2007xy}. Finally, we assume that (iv) the deceleration due to dipole magnetic breaking leads to a period evolution given by
\begin{equation}\label{eq:act_period}
P = \left(P_{0}^{2}+0.154\,B_{0}^{2}t\right)^{1/2}\,,
\end{equation}
where $P$ is in ms and $t$ is in Gyr. Our simulation gives us $N(f)$---the expected number density of neutron stars in the Galactic disk (per Hz) as a function of frequency.

Here we assume a birth rate of $5\times 10^{-4}$ {\em millisecond} neutron stars per century, corresponding to the upper estimate in~\cite{Story:2007xy}\footnote{The model of~\cite{Story:2007xy} includes a deathline in the plane $P$--$\dot{P}$ removing the subpopulation of ms pulsars not
observable in radio. In principle this selection effect do not apply to GW observations but pulsars above the deathline evolve very quickly toward periods outside of the detector frequency band so that including them would not change the results significantly.}.The model does not include the contribution from globular clusters.

\begin{figure}[h!]
{\includegraphics[scale = 0.4]{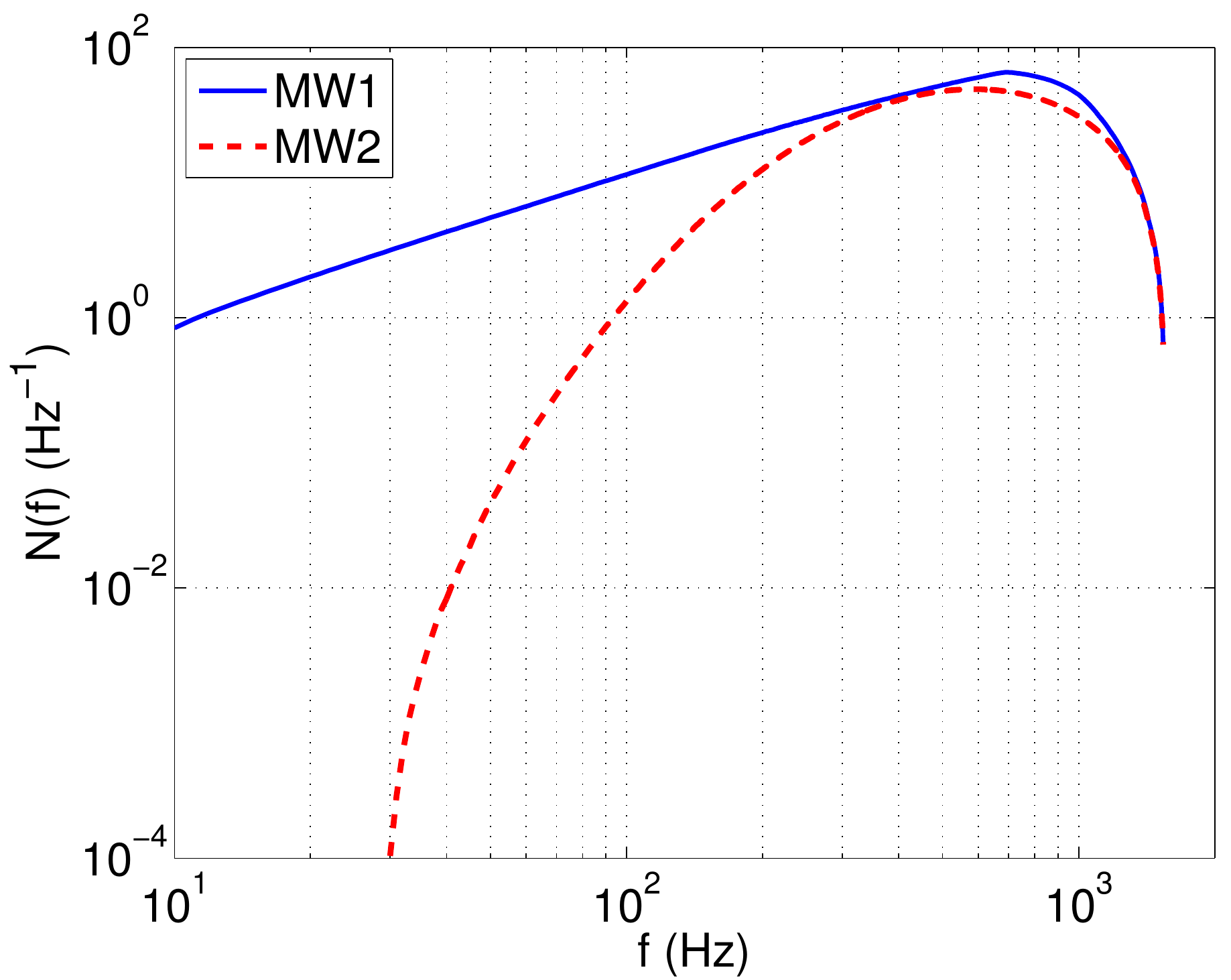}}
%  \psfig{file=Nf.eps,height=2.5in}
  % created by plot_N.m
  \caption{
    The number density of Milky Way neutron stars $N(f)$ as a function of gravitational-wave frequency.
%    The solid blue curve is the ``simple'' model while the dashed red is the Story et al. model.
    There are $N_\text{tot}=52800$ neutron stars in band ($10$--$1538\,\text{Hz}$) for MW1 and 40000 for MW2.
}
\label{fig:N_f_galaxy}
\end{figure}

The distribution of $N(f)$ is shown in Fig.~\ref{fig:N_f_galaxy} labeled by MW1. We also consider a model similar to MW1, where we assume a log-normal distribution of the initial magnetic field with mean $\langle\log(B_0)\rangle= 8.5$ and standard deviation 0.3. This distribution is consistent with the observed distribution listed in the Australia Telescope National Facility catalog~\cite{atnfcat}, which we expect
is not significantly affected by selection effects~\cite{Lorimer:private}. Its distribution of $N(f)$ is shown in Fig.~\ref{fig:N_f_galaxy} labeled by MW2.

The GW strain amplitude for each neutron star is given by~\cite{kramer}:
\begin{equation}
\label{eq:GWstrainfromNS}
  h_0(f) = 4\pi^2 \beta \frac{G \epsilon I}{c^4 r} f^2 \,,
\end{equation}
where $\beta(\leq 1)$ is the orientation factor~\cite{Dhurandhar:2011qe}, $I$ is its principal moment of inertia, $r$ is the distance to the source, $G$ is the Newton's gravitational constant, $c$ is the speed of light, and $\epsilon$ is the ellipticity.

\begin{figure}[h!]
  % created with plot_H.m.
{\includegraphics[scale = 0.4]{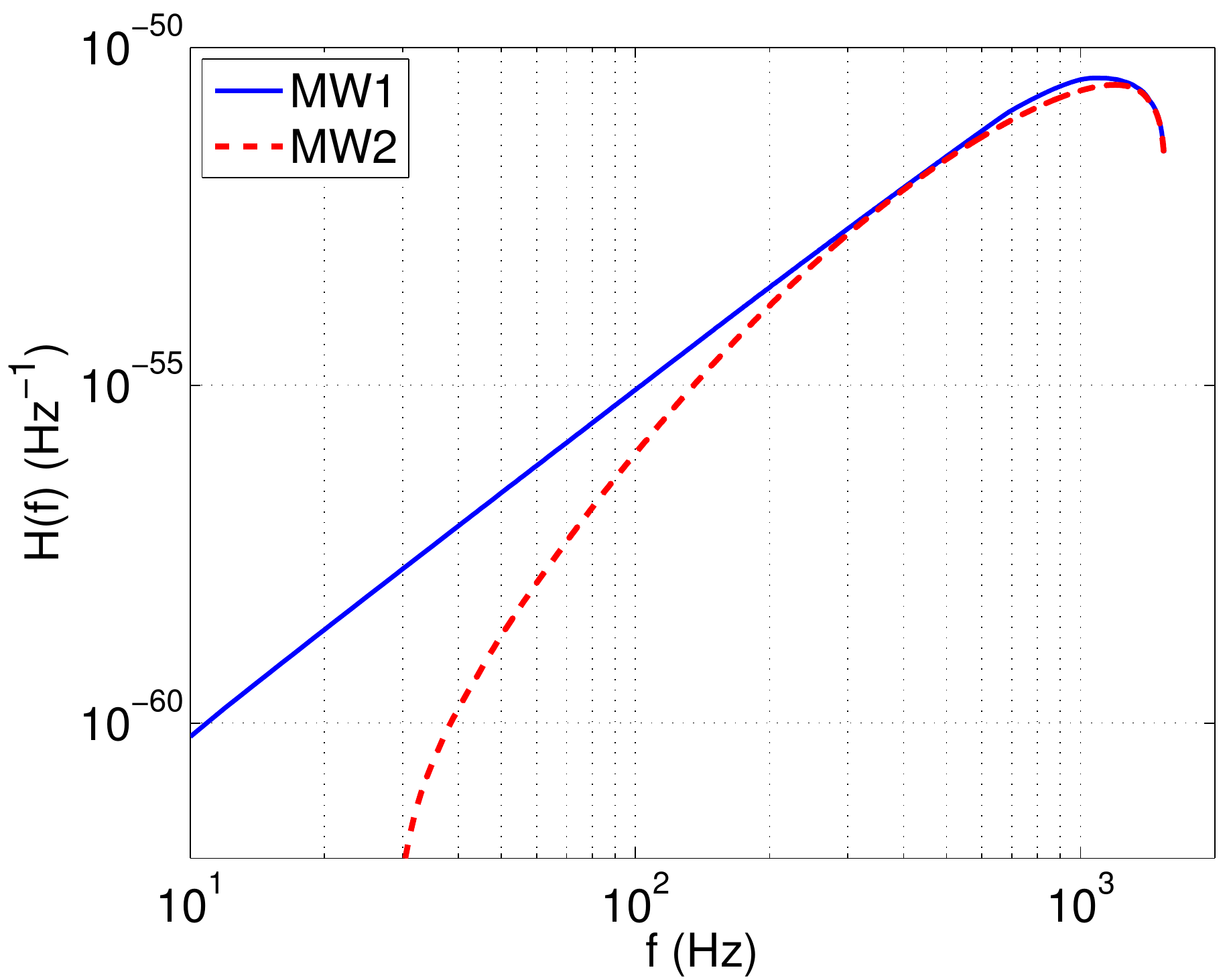}}
%  \psfig{file=Hf.eps,height=2.5in}
  % created by plot_H.m
  \caption{
    The power spectral density $H(f)$ of a stochastic signal formed from the incoherent superposition of gravitational waves from Galactic neutron stars.
    The solid blue dashed curve represents MW1 with $N_\text{tot}=52800$ while the dashed red represents MW2 with $N_\text{tot}=40000$.
    This plot is generated assuming $\langle\epsilon^{2}\rangle^{1/2}=10^{-7}$, the average distance measure $\langle 1/r^2 \rangle^{-1/2}=\unit[6.0]{kpc}$, and moment of inertia $I=\unit[1.1\times10^{45}]{g\, cm^{2}}$.
  }
\label{fig:Hf_plot_all}
\end{figure}

By combining Eq.~\eqref{eq:GWstrainfromNS} and the distribution of $N(f)$ from Fig.~\ref{fig:N_f_galaxy}, we can obtain the spectral shape of GW power spectral density $H(f)$ from Milky Way neutron stars:
\begin{equation}\label{eq:power0}
  \begin{split}
    H(f) = & \frac{1}{2} \langle h_0^*(f) h_0(f) \rangle \, N(f) \\
    = & 8 \pi^4 \frac{0.4 \, G^2 \langle\epsilon^2\rangle\langle I^2\rangle}
    {c^8} \left\langle\frac{1}{r^2}\right\rangle f^4 N(f) .
  \end{split}
\end{equation}
Here the angled brackets denote an expectation value.
The factor of $1/2$ comes from the fact that $h_0(f)$ is measured peak-to-peak whereas $H(f)^{1/2}$ is the root-mean-squared amplitude.
We have assumed that $\beta$, $I$, $\epsilon$, and $r$ are independent variables.
Also, we utilize the fact that $\langle \beta^2 \rangle = 0.4$, given the expected priors on neutron star inclination angle and polarization angle.
If we further assume that $I$, $\epsilon$, and $r$ are independent of frequency, then Eq.~\eqref{eq:power0} completely determines the shape of the GW power spectrum of the stochastic signal from Milky Way neutron stars. We plot $H(f)$ (with arbitrary normalization) in Fig.~\ref{fig:Hf_plot_all}.

Note that while $N(f)$ is peaked at $\approx\unit[600]{Hz}$, $H(f)$ is peaked at much higher frequencies due to the fact that $H(f)\propto f^4 N(f)$.
The normalization of $H(f)$ depends on the unknown normalization of $N(f)$ and the unknown average ellipticity.
Thus, in the analysis that follows, we rely on the {\em shape} of $H(f)$, but not the overall normalization.

With Fig.~\ref{fig:Hf_plot_all}, we have a working model for the distribution of neutron stars in frequency; we now consider their distribution in space.
% The following number comes from rsquared.m.
Using the distributions of neutron stars' radial distance from the Galactic center (and the vertical distance from the Galactic plane) from Ref.~\cite{Story:2007xy}, and assuming that the Earth is $\unit[8.3]{kpc}$ from the Galactic Center, we estimate that $\langle 1/r^2 \rangle^{-1/2} \approx \unit[6.0]{kpc}$.
We can thereby write Eq.~\eqref{eq:power0} as:
% pre-factor calculated with fac.m
\begin{widetext}
\begin{equation}\label{eq:power1}
  H(f) =
  \left(\unit[7.0\times10^{-27}]{Hz^{-1/2}}\right)^2
  \left(\frac{\langle\epsilon^2\rangle}{(1\times10^{-7})^2}\right)
  \left(\frac{\langle I^2 \rangle}{(\unit[1.1\times10^{45}]{g\,cm^2})^2}\right)
  \left(\frac{\langle 1/r^2 \rangle}{1/(\unit[6.0]{kpc})^2}\right)
  \left(\frac{f}{\unit[900]{Hz}}\right)^4
  N(f)
\end{equation}
\end{widetext}

For our present purposes, Eq.~\eqref{eq:power1} is nearly sufficient to describe the stochastic signal from Milky Way neutron stars.
For the sake of simplicity, in the discussion that follows we assume that $ \langle 1/r^2 \rangle$ and $\langle I^2 \rangle$ are reasonably well-constrained from population synthesis models and nuclear physics, respectively. This establishes a simple relationship 
\begin{equation}\label{eq:kappa}
H(f) = \kappa~\langle \epsilon^2 \rangle N(f)\,,
\end{equation}
where $\kappa$ is a proportionality factor that can be readily obtained from Eq.~(\ref{eq:power0}). By measuring $H(f)$ with GW detectors, it is therefore possible to constrain the product $\langle \epsilon^2 \rangle N(f)$.
This is the topic of the next section.
In Section~\ref{sec:summary}, we revisit our assumptions about $\langle 1/r^2 \rangle$ and $\langle I^2 \rangle$, and discuss how a more careful treatment can incorporate systematic error from our imperfect knowledge of these two quantities.

The final ingredient in a description of the stochastic signal from Galactic neutron stars is their angular distribution in the sky ${\cal P}(\hat{n})$.
A priori, we expect ${\cal P}(\hat{n})$ to be highly peaked toward the Galactic center.
Most previous searches for the stochastic background, however, assumed an isotropic distribution. For comparison, therefore, in the analysis that follows it will be useful to make the (inaccurate) assumption that ${\cal P} (\hat{n})$ is isotropic.

By assuming isotropy, our results will be overly conservative, though no less accurate.
This is because the sensitivity of detectors like LIGO to isotropic stochastic signals is diminished (in comparison to pointlike sources) by the signal interference encoded in the overlap reduction function~\cite{PhysRevD.59.102001,christensen_prd}.
The loss of signal due to the overlap reduction function affects only spatially separated detectors; colocated detectors are immune.
We therefore include results for both colocated and separated detectors. The results for colocated detectors limit the maximum possible improvement that can be achieved with more careful modeling of ${\cal P}(\hat{n})$.

\section{Methodology}\label{sec:methodology}
For the sake of simplicity, we assume a network of two detectors denoted $1$ and $2$.
As our starting point, we begin with an unbiased estimator for $H(f)$:
\begin{equation}\label{eq:pt_est}
  \widehat{Y}(f) = 
  \frac{1}{\cal N} \frac{5}{\gamma_{12}(f)}
  \text{Re} \left( \tilde{s}_1^*(f) \tilde{s}_2(f) \right) ,
\end{equation}
and the associated uncertainty:
\begin{equation}\label{eq:sigma}
  \widehat\sigma_Y(f) = \frac{1}{\sqrt{2}} \frac{5}{\gamma_{12}(f)} \sqrt{P_1(f) P_2(f)} .
\end{equation}
Here, $\tilde{s}_I(f)$ is the Fourier transform of the strain measured by detector $I$, ${\cal N}$ is a Fourier normalization constant, $P_I$ is the strain autopower spectrum for detector $I$, and $\gamma_{12}(f)$ is the normalized overlap reduction function for the detector pair~\cite{locus}.
The factor of $5$ comes from averaging the detector response over direction and polarization states.
$\widehat{Y}(f)$ and $\widehat\sigma_Y(f)$ are the standard outputs of isotropic stochastic analyses; see e.g., Ref.~\cite{Abbott:2009ws}. The wide-hat on a quantity denotes its estimator. $\widehat{Y}(f)$ can be rewritten in units of energy density:
\begin{equation}
  \widehat\Omega(f) = \frac{2\pi^2}{3 H_0^2} f^3 \widehat{Y}(f) ,
\end{equation}
where $H_0$ is the Hubble constant.
In this paper we take $H_0 =\unit[68]{km\,sec^{-1}Mpc^{-1}}$ \cite{Ade:2013zuv}.

From Eq.~\eqref{eq:pt_est} and Eq.~\eqref{eq:power1} we can obtain the following estimators for average neutron star ellipticity squared, given $N(f)$:
\begin{widetext}
\begin{equation}\label{eq:epsilon}
  \begin{split}
  \widehat{\epsilon^2}(f) = & 
  \left(1\times10^{-7}\right)^2
  \left[
  \frac{1}{N(f)}
  % the prefactor is determined by squaring the prefactor in the expression for
  % H(f)
  \left(\frac{\widehat{Y}(f)}{\unit[4.9\times10^{-53}]{Hz^{-1}}}\right)
  \left(\frac{(\unit[1.1\times10^{45}]{g\, cm^2})^2}{\langle I^2 \rangle}\right)
%  \left(\frac{\langle r^2 \rangle}{(\unit[14.9]{kpc})^2}\right)
%
  \left(\frac{1/(\unit[6.0]{kpc})^2}{\langle 1/r^2 \rangle}\right)
  \left(\frac{\unit[900]{Hz}}{f}\right)^4
  \right] \\ 
  = &
  \left(1\times10^{-7}\right)^2
  \left[
  \frac{1}{N(f)}
  \left(\frac{\widehat{\Omega}(f)}{4.8\times10^{-8}}\right)
  \left(\frac{(\unit[1.1\times10^{45}]{g\, cm^2})^2}{\langle I^2 \rangle}\right)
%  \left(\frac{\langle r^2 \rangle}{(\unit[14.9]{kpc})^2}\right)
  \left(\frac{1/(\unit[6.0]{kpc})^2}{\langle 1/r^2 \rangle}\right)
  \left(\frac{\unit[900]{Hz}}{f}\right)^7
  \right] .
  \end{split}
\end{equation}
\end{widetext}

Eq.~\eqref{eq:epsilon} is framed in terms of ellipticity squared, but it is more convenient to work with just ellipticity.
We can write the expectation value of $\widehat{\epsilon^2}(f)$ as
\begin{equation}
  \langle\widehat{\epsilon^2}(f)\rangle = 
  \epsilon^2(f) + \Sigma_\epsilon^2(f) ,
\end{equation}
where $\Sigma_\epsilon^2(f)$ is the intrinsic variance of the ellipticity distribution and $\epsilon(f)$ is the mean value.
(We use capital $\Sigma_\epsilon^2(f)$ to denote the intrinsic variance and lower-case $\sigma_\epsilon^2(f)$ to denote the variance associated with the estimator $\widehat\epsilon(f)$ defined in Eq.~\eqref{eq:epsilonhat}.)
Physical ellipticity is a positive definite quantity.
Thus, it is possible to make the rough approximation that
\begin{equation}
  \langle\widehat{\epsilon^2}(f)\rangle \approx \epsilon^2(f) .
\end{equation}
This is an excellent approximation if, for example, ellipticity turns out to be log-normally distributed.
If, on the other hand, ellipticity is exponentially distributed, then $\Sigma_\epsilon(f)=\epsilon(f)$, but even then, the approximation results in a modest $40\%$ overestimate of $\epsilon(f)$.

Thus, for the sake of simplicity, we define the following {\em biased} estimator for average ellipticity:
\begin{equation}\label{eq:epsilonhat}
  \widehat\epsilon(f) = \sqrt{\widehat{\epsilon^2}(f)} .
\end{equation}
Since ellipticity is positive-definite, there is good motivation for supposing that the bias associated with $\widehat\epsilon(f)$ is relatively small, and so this approximation will be a useful simplifying assumption.
Moreover, sensitivity estimates derived with $\widehat\epsilon(f)$ will be conservative since non-zero $\Sigma_\epsilon(f)$ will tend to increase the detectability of a stochastic signal given a fixed $\epsilon(f)$.

We henceforth work with $\widehat\epsilon(f)$ under the assumption that $\Sigma_\epsilon(f)\lesssim\epsilon(f)$.
In the event that a stochastic signal from Galactic neutron stars is detected, there are at least two ways to potentially account for the bias.
First, $\Sigma_\epsilon(f)$ could be estimated using measurements of individual neutron stars.
Second, $\Sigma_\epsilon(f)$ could be estimated using a theoretical model.

%Note that the expectation value of $\widehat\epsilon$ will have a (probably small) bias toward larger values due to the intrinsic width of the ellipticity distributions $\sigma_\epsilon$:
%\begin{equation}
%  \langle\widehat\epsilon(f)\rangle =
%  \sqrt{\epsilon^2(f) + \sigma_\epsilon^2(f)} .
%\end{equation}
It is worthwhile to note how $\widehat\epsilon(f)$ depends on other parameters.
We obtain more constraining limits [$\epsilon(f)$ is smaller] when $N(f)$ is increased (we assume the existence of more neutron stars) and when $\widehat{\sigma}_Y(f)$ is decreased (the detector is less noisy).

The uncertainty associated with $\widehat\epsilon(f)$ (Eq.~\eqref{eq:epsilonhat})---denoted $\sigma_\epsilon(f)$---can be expressed in terms of $\sigma_\Omega(f)$ [or, equivalently, $\widehat\sigma_Y(f)$] as follows.
The likelihood functions for $\widehat{Y}(f)$ and $\widehat\Omega(f)$ are known to be essentially Gaussian~\cite{PhysRevD.59.102001,Abbott:2009ws,PhysRevD.85.122001}, e.g.,:
\begin{equation}\label{p_Omega}
  p_\Omega(\widehat\Omega(f)|\Omega(f)) = 
  \frac{1}{\sqrt{2\pi}\sigma_\Omega(f)}e^{-(\Omega(f)-\widehat\Omega(f))^2/2\sigma_\Omega^2(f)} .
\end{equation}
It follows from Eqs.~(\ref{eq:epsilon}) and (\ref{eq:kappa}) that the likelihood function for $\widehat\epsilon(f)$ is given by
\begin{equation}\label{p_epsilon}
  p_\epsilon(\widehat\epsilon(f)|\epsilon(f)) = 
  \sqrt{\frac{8}{\pi}}\,\frac{\epsilon(f)}{\kappa \sigma_\Omega(f)}
  \,e^{-(\epsilon^2(f)-\widehat\epsilon^2(f))^2/2\kappa^2 \sigma_\Omega^2(f)} \,,
\end{equation}
which is not a Gaussian distribution.
The latter function, however, is simple enough such that its mean and variance can be obtained in closed form in some special cases.
One such case is when $\widehat\Omega(f)=\widehat\epsilon(f)=0$.
In that event, the mean and variance of the distribution in Eq.~\eqref{p_epsilon} are
\begin{equation}\label{eq:mu_and_sigma}
  \begin{split}
    \left[ \langle \epsilon(f) \rangle \right]_{\widehat\epsilon(f) = 0}
    = & \frac{2^{3/4}\sqrt{\pi\kappa \sigma_\Omega(f)}}{\Gamma\left(\frac{1}{4}\right)} \\
    \approx & 0.82 \sqrt{\kappa\sigma_\Omega(f)} \\
    \sigma^2_\epsilon(f)\Big|_{\widehat\epsilon(f) = 0} = &
    \left[ \langle \epsilon^2(f) \rangle -
      \langle \epsilon(f) \rangle^2 \right]_{\widehat\epsilon(f) = 0} \\
    = & \left[\sqrt{\frac{2}{\pi}} - \frac{2^{3/2}\pi}{\Gamma^2\left(\frac{1}{4}\right)}\right]\,\kappa \sigma_\Omega(f) \\
    \approx & 0.12\kappa \sigma_\Omega(f) ,
  \end{split}
\end{equation}
assuming that physical values of $\Omega(f)$ and $\epsilon(f)$ must be positive.

In the more desirable case of $\widehat\epsilon(f)> 0$, one finds that
\begin{equation}
  \begin{split}
%  \mu_\epsilon(f) [\widehat\epsilon(f) \neq 0] 
  \langle \epsilon(f) \rangle
  =& \frac{\sqrt{\kappa \sigma_\Omega(f)}\,e^{-\widehat\epsilon^4(f)/(4\kappa^2\sigma_\Omega^2(f))}}{\sqrt{2}} \\  &
  \times D_{-3/2}\left(-\frac{\widehat\epsilon^2(f)}{\kappa \sigma_\Omega(f)}\right)\,,
 \end{split}
\end{equation}
where $D_{-3/2}$ is a parabolic cylinder function. 
It is straightforwardly shown that this expression yields the correct value in the limit of vanishing $\widehat\epsilon(f)$.

Searches for the stochastic background gain a significant boost in sensitivity through the optimal combination of measurements from many frequency bins~\cite{locus}.
Using this principle, and assuming $\epsilon$ is independent of frequency, we obtain an optimal broadband estimator:
\begin{equation}\label{eq:epsilon_opt}
  \widehat\epsilon_\text{opt} = 
  \frac{\sum_{f} \widehat\epsilon(f) \, \widehat\sigma_\epsilon^{-2}(f)}
  {\sum_f \widehat\sigma_\epsilon^{-2}(f)} ,
\end{equation}
with associated uncertainty
\begin{equation}
  \widehat\sigma_\text{opt} = 
  \left(\sum_f \widehat\sigma_\epsilon^{-2}(f)\right)^{-1/2} .
\end{equation}
Note that $\widehat\sigma_\epsilon(f)$ is an estimator for the uncertainty associated with $\widehat\epsilon(f)$ whereas $\Sigma_\epsilon(f)$ is the intrinsic width of the distribution of $\epsilon(f)$.

In order to calculate $\widehat\epsilon_\text{opt}$, we need to know the shape of $N(f)$.
This allows us to weight different frequency bins based on the expected number of neutron stars in each bin.
However, the absolute normalization of $N(f)$ is unknown.
Since $\widehat\epsilon_\text{opt}\propto N^{-1/2}(f)$, the product $\widehat\epsilon_\text{opt}^2 N_\text{tot}$ does not depend on the overall normalization of $N(f)$.
Thus, to minimize systematic errors from theoretical unknowns, it is useful to constrain the quantity $\epsilon^2 N_\text{tot}$ where
\begin{equation}
  N_\text{tot} \equiv \int_\text{band} \mathrm{d}f \, N(f)
\end{equation}
is the total number of neutron stars emitting in some observing band.
In the next section we apply this formalism to constrain $\epsilon^2 N_\text{tot}$ using previously published results.
We also estimate the sensitivity of future possible observations.

\section{Results}\label{sec:searchresults}
\subsection{Projected sensitivity of current and planned observatories}
\begin{figure}[h!]
{\includegraphics[scale = 0.4]{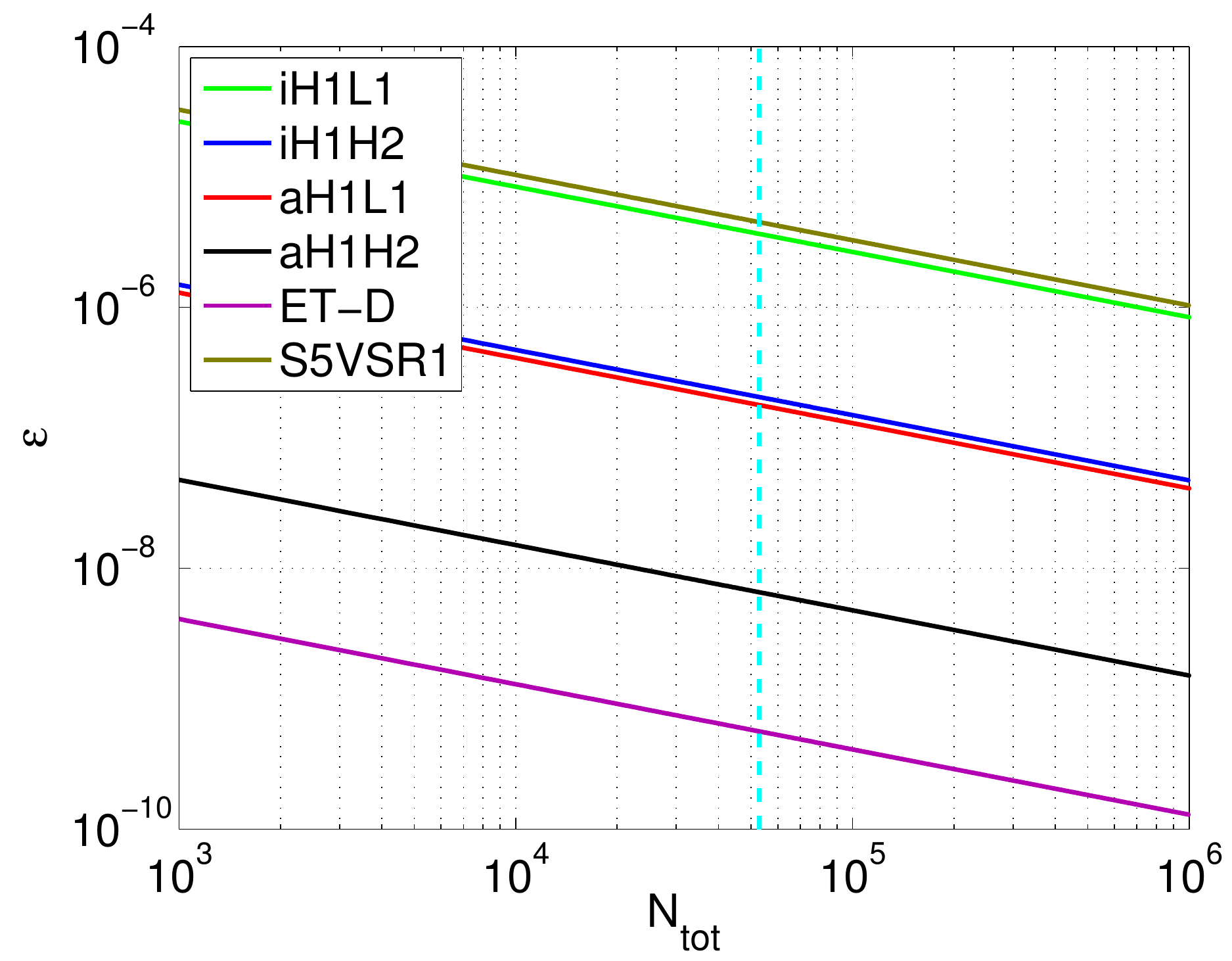}}
%  \psfig{file=cal_contours.eps,height=2.5in}
  % created by cal_epsilon4.m
  \caption{
    One-sigma sensitivity curves in the $\epsilon$--$N_\text{tot}$ plane.
    The parameter space above a curve will, on average, produce a signal with a signal-to-noise ratio greater than unity.
    We show theoretical sensitivity curves (assuming  $\unit[1]{yr}$ of observation time) for initial LIGO H1L1, initial LIGO H1H2, Advanced LIGO H1L1~\cite{aligo}, Advanced LIGO H1H2, and for the Einstein Telescope.
    (While efforts are underway to move H2 to India, we include the H1H2 detector pair for illustrative purposes.)
    We also show the measured sensitivity obtained in previously published results using data from both initial LIGO and Virgo~\cite{PhysRevD.85.122001}.
    The vertical dashed cyan line indicates $N_\text{tot}=52800$.
    These results are obtained using MW1.
    Results obtained with MW2 agree to within $15\%$ in $\epsilon$.
  }
  \label{fig:esquared_N}
\end{figure}

In Fig.~\ref{fig:esquared_N} we present the projected one-sigma sensitivity for a variety of experiments in the $\epsilon$--$N_\text{tot}$ plane assuming $\unit[1]{yr}$ of observation time.
We include projections for initial LIGO and Advanced LIGO using the spatially separated H1L1 detector network and the colocated H1H2 detector pair. Here we use publicly available sensitivity curves~\cite{iLigoCurves,aLigoCurves}.
Work is underway to relocate the H2 detector to India for Advanced LIGO, but we include the colocated pair to make comparisons with projections for the Einstein Telescope~\cite{0264-9381-28-9-094013}, which has colocated interferometers in its design.
%We also include a projection for the Einstein Telescope~\cite{0264-9381-28-9-094013} and 
We also include the sensitivity obtained from a previously published analysis by initial LIGO and Virgo~\cite{PhysRevD.85.122001}.

\begin{figure}[h!]
{\includegraphics[scale = 0.4]{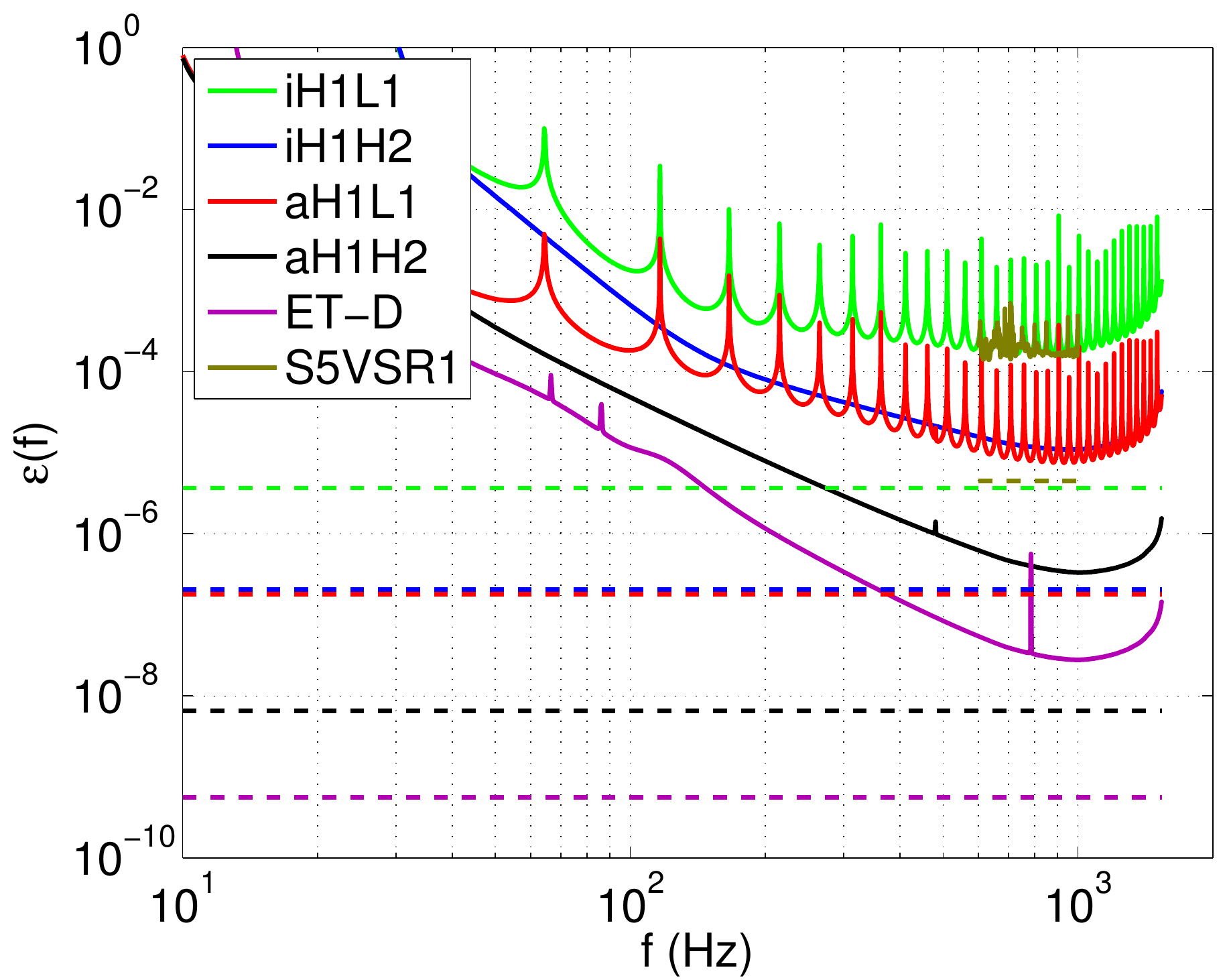}}
%  \psfig{file=epsilon.eps,height=2.5in}
  % created by cal_epsilon4.m
  \caption{
    One-sigma sensitivity to average ellipticity of Galactic neutron stars. The solid lines are narrowband results calculated with $\unit[0.25]{Hz}$-wide bins whereas the dashed lines show the broadband values of $\sigma_\text{opt}$ obtained through the optimal combination of all the frequency bins. This plot assumes $N_\text{tot}=52800$ in-band neutron stars, average distance squared $\langle 1/r^2 \rangle^{-1/2}=\unit[6.0]{kpc}$, and moment of inertia $I=\unit[1.1\times10^{45}]{g\, cm^{2}}$. We show theoretical sensitivity curves (assuming $\unit[1]{yr}$ of observation time) for initial LIGO H1L1, initial LIGO H1H2, Advanced LIGO H1L1, Advanced LIGO H1H2, and for the Einstein Telescope.
    The difference in spectral shape between the S5VSR1 curve and the iH1L1 curve is due to the different overlap reduction functions for H1L1, H1V1, and L1V1.    (While efforts are underway to move H2 to India, we include the H1H2 detector pair for illustrative purposes.)
    We also show the measured sensitivity obtained in previously published results using data from both initial LIGO and Virgo~\cite{PhysRevD.85.122001}.
    These results are obtained using model MW1.
    Results obtained with model MW2 agree to within $15\%$ in $\epsilon$ for $N_\text{tot}=40000$.
  }
  \label{fig:eps_spec}
\end{figure}

In Fig.~\ref{fig:eps_spec} we consider the case where $N_\text{tot}=52800$ in order to see how well $\epsilon$ can be constrained by stochastic measurements given a plausible value of $N_\text{tot}$. 
The solid curves show the sensitivity as a function of frequency whereas the dashed curves show the combined broadband sensitivity $\sigma_\text{opt}$.
The values of $\sigma_\text{opt}$ are summarized in Table~\ref{tab:sigma}.

For initial LIGO, it might be possible to achieve $\sigma_\text{opt}\approx2\times10^{-7}$ using the colocated H1H2 detector pair.
This is also close to what can be achieved during Advanced LIGO with the H1L1 detector network.
A pair of colocated Advanced LIGO detectors could, in principle, achieve a sensitivity of $\approx7\times10^{-9}$, which is an order of magnitude better than the current limits on individual neutron star ellipticity from targeted GW searches~\cite{S6-known}.
As we pointed out above, a more sophisticated analysis, with an improved model for the anisotropy of the stochastic signal, will likely yield a sensitivity for aH1L1 that is somewhere between the aH1L1 and aH1H2 sensitivities given in Table~\ref{tab:sigma}.

The inclusion of additional detectors such as Virgo and KAGRA~\cite{Kagra} is expected to improve the results marginally since the overlap reduction function is most favorable for the LIGO pair, though, this remains an area of future investigation.
Finally, the proposed Einstein Telescope is expected to achieve a sensitivity of $\sigma_\text{opt}\approx6\times10^{-10}$. 
This is significantly below the current best limits on neutron star ellipticity~\cite{S6-known}, which suggests that the Einstein Telescope may have sufficient sensitivity to observe a stochastic signal from Galactic neutron stars.

\begin{table}
  \begin{tabular}{|c|c|c|}
    \hline
    network & $\sigma_\text{opt}$ \\\hline
    iH1L1 & $3.7\times10^{-6}$ \\\hline
    iH1H2 & $2.1\times10^{-7}$ \\\hline
    aH1L1 & $1.8\times10^{-7}$ \\\hline
    aH1H2 & $6.7\times10^{-9}$ \\\hline    
    ET-D &  $5.6\times10^{-10}$ \\\hline
    S5VSR1 & $4.5\times10^{-6}$ \\\hline
    % calculated with cal_epsilon4.m
  \end{tabular}
  \caption{
    One-sigma sensitivity to ellipticity from Galactic neutron stars assuming the model MW1 (with $N_\text{tot}=52800$ in-band neutron stars), average distance squared $\langle 1/r^2 \rangle^{-1/2}=\unit[6.0]{kpc}$, and moment of inertia $I=\unit[1.1\times10^{45}]{g\, cm^{2}}$.
    (Results obtained with model MW2 agree to within $15\%$.)
    We assume a cut-off frequency $f\leq\unit[1538]{Hz}$. Results are shown for  initial LIGO H1L1, initial LIGO H1H2, Advanced LIGO H1L1, Advanced LIGO H1H2, and for the Einstein Telescope.
    Each entry is calculated assuming $\unit[1]{yr}$ of integration except for S5VSR1, which is derived from a  previously published paper from initial LIGO and Virgo~\cite{PhysRevD.85.122001}.
  }
  \label{tab:sigma}
\end{table}

\subsection{Combining with measurements of resolvable neutron star signals}

It is interesting to consider what we can learn by combining a stochastic background measurement with GW measurements of individual neutron stars.
Since the latter constrain ellipticity directly, searches for individual neutron stars can be used to break the degeneracy in stochastic background measurements between $\epsilon$ and $N_\text{tot}$, allowing us to estimate the number of in-band Galactic neutron stars.
In this subsection we estimate roughly how well we can constrain $N_\text{tot}$ by combining a stochastic search with a future GW detection of individual neutron stars.

To begin, we assume that the fractional uncertainty in $\epsilon$ from measurements of individual neutron stars is small compared to the fractional uncertainty in $\Omega\propto\epsilon^2 N_\text{tot}$, which is measured by a stochastic search. It follows that the fractional uncertainty on $N_\text{tot}$ is $\sigma_N/N_\text{tot}\approx\sigma_\Omega/\Omega\approx1/\text{SNR}$. If we imagine, for example, that the Einstein Telescope is able to detect an $\text{SNR}=5$ stochastic signal from Galactic neutron stars, and that the average $\epsilon$ is by then tightly constrained from observations of individual neutron stars, it should be possible to estimate $N_\text{tot}$ to within a single-sigma uncertainty of $\approx20\%$.

\section{Conclusions}\label{sec:summary}
We have shown how observations of stochastic gravitational waves can be used to constrain both the number of Galactic neutron stars in some analysis band $N_\text{tot}$ as well as their average ellipticity $\epsilon$.
We calculate the sensitivity of past, present, and future experiments in the $\epsilon$--$N_\text{tot}$ plane.
We demonstrate that our predictions are fairly robust to details in the modeling of Galactic neutron stars.

For the reasonable values of $N_\text{tot}\approx40000$--$53000$, we find that a colocated pair of initial LIGO detectors can, in principle, achieve a sensitivity of $\sigma_\epsilon\approx2\times10^{-7}$, which is already an interesting part of parameter space.
Advanced LIGO, without a colocated detector pair, may have difficulty improving significantly on the sensitivity of a colocated initial LIGO pair.
However, the proposed Einstein Telescope will be able to probe $\sigma_\epsilon\approx6\times10^{-10}$.
We demonstrate that stochastic measurements can be combined with measurements of individually resolvable neutron star signals in order to break the degeneracy between $\epsilon$ and $N_\text{tot}$, thereby providing an estimate of the total number of Galactic neutron stars in band.

A promising area of future work is the development of directional Galactic search for stochastic gravitational waves.
Using a $\lambda$-statistic analysis (as in Appendix~\ref{sec:virgoresults}), it should be possible to improve the sensitivity (for non-colocated detectors) beyond the estimates stated here. A directional analysis---combined with measurements of individual neutron stars---might also provide further information, e.g., about the spatial distribution of neutron stars in the Milky Way. The analysis can be further improved by taking into account theoretical uncertainty in the expectation values $\langle I^2 \rangle$ and $\langle 1/r^2 \rangle$, which are used in the estimation of $\epsilon^2 N_\text{tot}$.

\acknowledgments

We thank Nelson Christensen, Peter Gonthier, Duncan Lorimer, Matthew Pitkin, Keith Riles, and Graham Woan for helpful discussions and comments. We gratefully acknowledge National Science Foundation for funding LIGO, and the LIGO Scientific Collaboration and the Virgo Collaboration for access to this data. This work is supported in part by NSF Grants No. PHY-1205952, PHY-1206108, and PHY-1307401. ET is a member of the LIGO Laboratory, supported by funding from United States National Science Foundation. LIGO was constructed by the California Institute of Technology and Massachusetts Institute of Technology with funding from the National Science Foundation and operates under cooperative agreement PHY-0757058.

\appendix
\section{Stochastic background from the Milky Way, the Virgo Cluster, and the entire universe}

In this paper we have derived constraints on the average properties of Milky Way neutron stars by considering their combined stochastic gravitational-wave signal.
It is worthwhile to consider if this is, indeed, the best means of constraining average properties of neutron stars.
While the Milky Way contains $\approx40000$ neutron stars in the band of Advanced LIGO, the $\approx1000$ galaxies making up the Virgo Cluster contain many more.
While the Virgo Cluster contains many more neutron stars, they are further away.
A typical Galactic distance is $\unit[10]{kpc}$ whereas the the Virgo Cluster is significantly further away $\approx\unit[16.5]{Mpc}$.
Which source produces a brighter gravitational-wave signal: the nearby neutron stars of the Milky Way or the more distant, but more numerous neutron stars of the Virgo Cluster?
For that matter, how do these two signals compare to the signal arising from the extremely large number of neutron stars in the entire universe, the vast majority of which are very far away?
These questions, which we attempt to answer here, amount to a variation on Olbers' paradox (see Ref~\cite{Mazumder:2014fja} for further discussions).

Our answer consists of a back-of-the-envelope calculation.
We begin by comparing the signal from the Milky Way with the signal from the Virgo Cluster.
Assuming a network of two identical Advanced LIGO detectors operating at design sensitivity with strain noise power spectral density $P(f)$, the expected signal-to-noise ratio from a stochastic neutron star signal~\cite{locus} scales like
\begin{equation}\label{eq:snr_scaling0}
  \text{SNR} \propto \left[ \int \mathrm{d}f \frac{\gamma^2(f)H^2(f)}{P^2(f)} \right]^{1/2} .
\end{equation}
Combining Eq.~\eqref{eq:snr_scaling0} with Eq.~\eqref{eq:power1},
\begin{equation}\label{eq:snr_scaling}
  \text{SNR} \propto \left\langle\frac{1}{r^2}\right\rangle \left[
    \int \mathrm{d}f \frac{\gamma^2(f) f^8 N^2(f)}{P^2(f)}
    \right]^{1/2} .
\end{equation}
Here $r$ is the distance to the neutron stars, $N(f)$ is the number of neutron stars in given frequency bin, $P(f)$ is the strain power spectral density of the detectors (assumed to be identical), and $\gamma(f)$ is the overlap reduction function.

The factor of $\langle 1/r^2 \rangle$ encodes the advantage of looking at nearby sources whereas the factor of $N(f)$ describes the advantage gained by looking at a source with more neutron stars.
The overlap reduction $\gamma(f)$ penalizes searches for diffuse sources, which create less easily detectable signal than pointlike sources.
The factor of $f^8$ arises through Eq.~\eqref{eq:GWstrainfromNS}.

Plugging in $\langle 1/r^2\rangle^{-1/2}=\unit[6]{kpc}$ for the Milky Way and and $\langle 1/r^2\rangle^{-1/2}=\unit[16.5]{Mpc}$ for the Virgo Cluster, and assuming $N(f)$ is 1000 times larger for the Virgo Cluster, we evaluate Eq.~\eqref{eq:snr_scaling} with the Advanced LIGO noise curve.
Using the Hanford-Livingston detector pair, we find that the SNR from the Milky Way is $\approx67\times$ greater than that from the Virgo Cluster.
Using colocated detectors, the Milky Way SNR is $\approx140\times$ greater due to the more favorable overlap reduction function for colocated detectors.

\section{Measuring a stochastic background from the Virgo Cluster}
\label{sec:virgoresults}
In this section we present a framework for measuring a stochastic signal from a population of neutron stars in the Virgo Cluster. As we demonstrated above, the Virgo Cluster search is expected to yield a less stringent constraint on neutron star ellipticity than the Milky Way search, which is the focus of this paper.
However, we include this example to demonstrate a general framework for measuring neutron star ellipticity with an anisotropic stochastic background.
We expect this demonstration to be useful for future work targeting an anisotropic Milky Way source.

The angular extent of the Virgo Cluster is about $6^\circ$ in radius, hence we consider it a localized source in the stochastic GW searches.
We apply multibaseline GW radiometry, a method that is optimal for searching for a localized stochastic signal with a network of detectors~\cite{PhysRevD.83.063002}.

The search statistic itself is derived from the cross correlations of the data across all possible baselines in the network. Following Refs.~\cite{PhysRevLett.107.271102,PhysRevD.80.122002} the GW energy density is characterized by 
\begin{equation}\label{eq:GWspectrum}
\Omega(f) =\frac{2\pi^{2}}{3H_{0}^{2}}f^{3}\bar{H}(f)\int_{S^{2}}\mathrm{d}\hat{n}\,{\mathcal{P}}(\hat{n})\,,
\end{equation} 
where ${\mathcal{P}}(\hat{n})$ specifies the angular distribution of GW power as a function of the sky-position unit vector $\hat{n}$, and $\bar{H}(f)$ its spectral shape. Note that $\bar{H}$, defined as $\bar{H}(f)\equiv H(f)/H(f_{0})$, is a dimensionless function of frequency, normalized so that $\bar{H}(f_{0})=1$, where $f_{0}$ is a reference frequency. An extended source with an arbitrary angular distribution can be expanded in spherical harmonic basis as
\begin{equation}\label{eq:sph_decomposition}
{\cal P}(\hat{n})=\sum_{l,m}{\cal P}^{lm}Y_{lm}(\hat{n})\,.
\end{equation}
The series is truncated at $l=l_{\text{max}}$, which sets the angular scale of the search to be $\sim 2\pi/l_{\text{max}}$. The choice of $l_{\text{max}}$ is determined by the detector network's angular resolution and the source power spectrum. 

\begin{figure}[h!]
{\includegraphics[scale = 0.4]{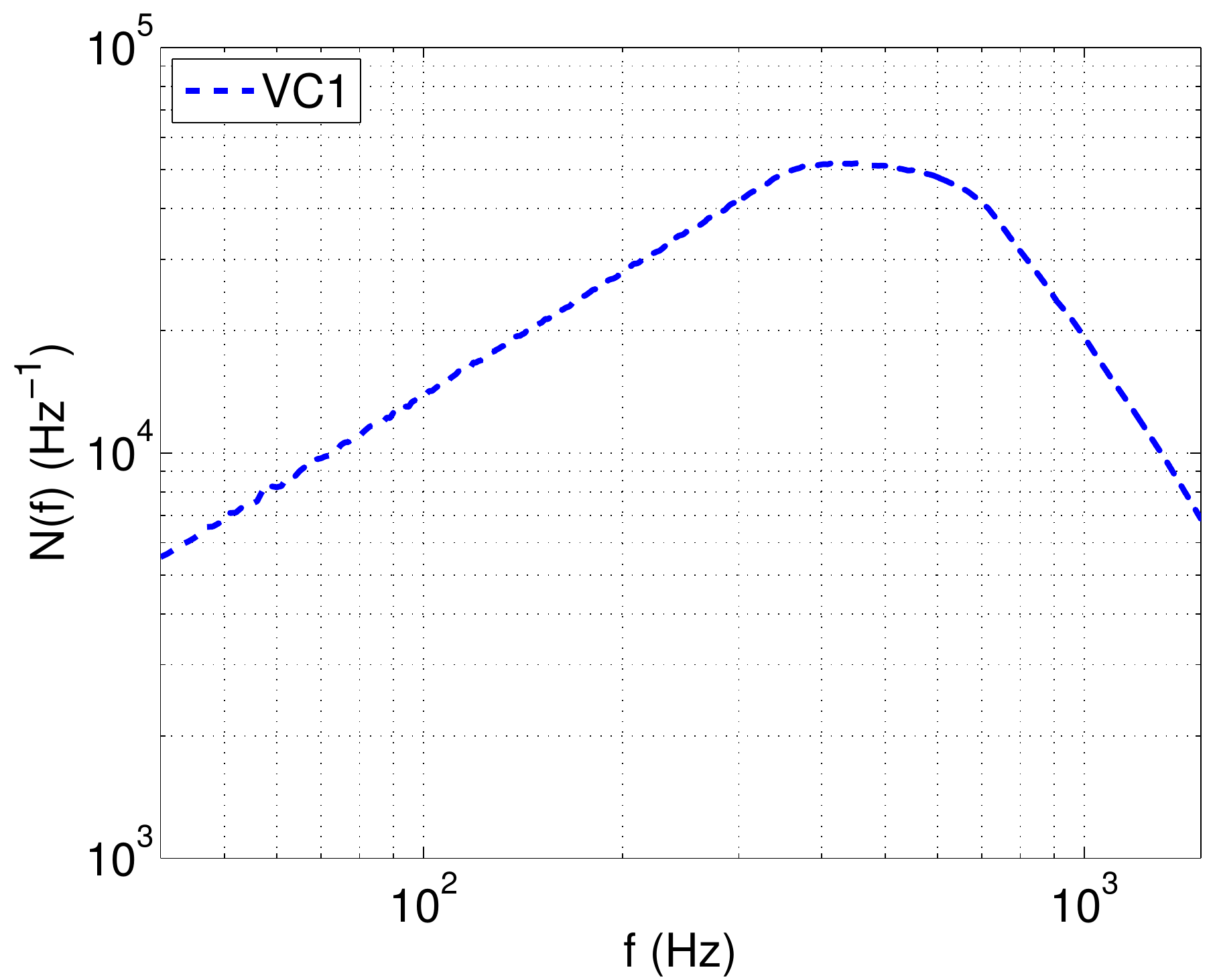}}
  \caption{
    The number density of Virgo Cluster neutron stars $N(f)$ as a function of gravitational-wave frequency.
    For this model, the normalization is chosen here to be $N_\text{tot}=4\times 10^{7}$ neutron stars in the analysis band of $\unit[40]{Hz}$--$\unit[1500]{Hz}$.
}
\label{fig:N_f_virgo}
\end{figure}

Following Ref.~\cite{PhysRevD.83.063002} one can combine the information about the network geometry and the source to define the following multibaseline statistic for detecting a stochastic signal:
\begin{equation}\label{eq:lambda_stat}
\lambda = \frac{\hat{\cal P}^{\mu\dagger}X_{\mu}}{\sqrt{\hat{\cal P}^{\nu\dagger}\Gamma_{\nu\tau}{\hat{\cal P}}^{\tau}}}\,,
\end{equation}
where $\mu\equiv \{l, m\}$, ${\hat{\bm{\mathcal{P}}}}$ is the unit vector along ${\bm{\mathcal{P}}}$, $X_{\nu}$ is the dirty map, and $\Gamma_{\mu\nu}$ is the beam matrix or Fisher matrix defined in Ref.~\cite{PhysRevLett.107.271102}. The maximized-likelihood ratio statistic in Eq.~\eqref{eq:lambda_stat} is obtained by maximizing the likelihood ratio over the overall source power $\alpha$, which is defined by ${\bm{\mathcal{P}}}\equiv \alpha {\hat{\bm{\mathcal{P}}}}$. The estimator of the overall power is given by
\begin{equation}\label{eq:estimator}
\widehat{\alpha} = \frac{\hat{\cal P}^{\mu\dagger}X_{\mu}}{\hat{\cal P}^{\nu\dagger}\Gamma_{\nu\tau}{\hat{\cal P}}^{\tau}}\,,
\end{equation}
with the associated uncertainty 
\begin{equation}\label{eq:uncertainty}
\sigma_{\widehat{\alpha}}=\frac{1}{\sqrt{\hat{\cal P}^{\nu\dagger}\Gamma_{\nu\tau}{\hat{\cal P}}^{\tau}}}\,.
\end{equation}
Note that $\widehat{\alpha}$ has units of $\mbox{strain}^{2}\mbox{Hz}^{-1}\mbox{sr}^{-1}$. One can extend these single-baseline quantities to the case of a multibaseline network, which is discussed in Ref.~\cite{PhysRevD.83.063002}. 

\begin{figure}
{\includegraphics[scale = 0.52]{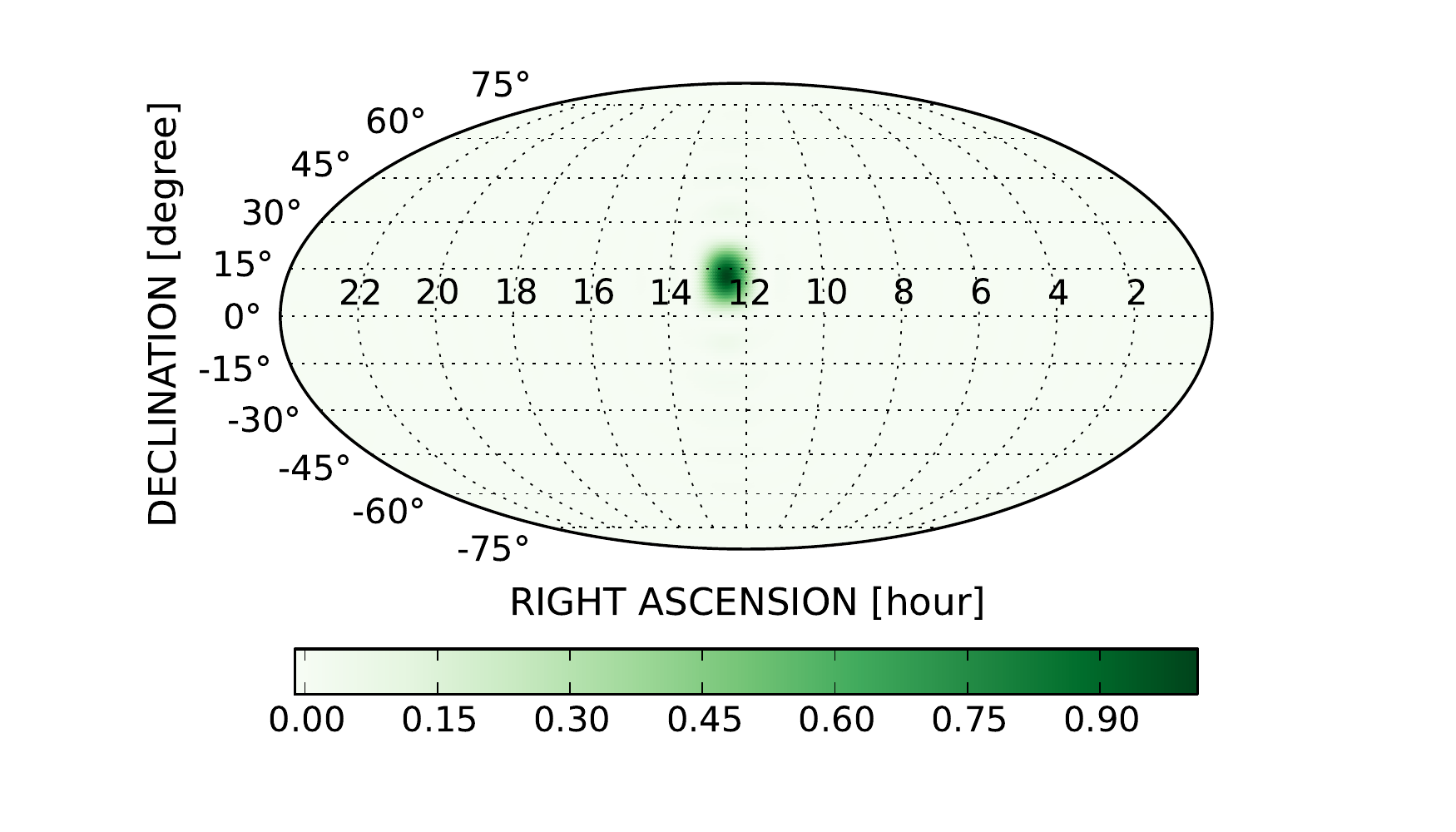}}
\caption{The map of a modeled signal-strength vector of the Virgo Cluster. In this map, the spherical harmonic coefficients are set to zero for $l > (l_{\text{max}} = 30)$.}
\label{fig:maps}
\end{figure}

We model the signal-strength vector $\hat{\bm{\mathcal{P}}}_{\text{model}}$ of the Virgo Cluster such that its non-zero components follow a Gaussian distribution, centered at 12h 26m 32s RA and $+12^{\circ}43^{'}24^{''}$ Dec, extended over 12h-13h RA and $5^{\circ}-20^{\circ}$ Dec. Figure~\ref{fig:maps} shows the signal-strength vector for the Virgo Cluster. The spherical-harmonic coefficients ${\mathcal{P}}^{lm}$ are set to zero for $l > (l_{\text{max}} = 30)$. This corresponds to an angular scale of about 12 degrees.

In order to estimate the sensitivity of our searches to the signal from the Virgo Cluster, we analyze data from the LIGO fifth science and the Virgo first science runs~\cite{iligo, Virgoupdate}. Here we use LIGO and Virgo time-shifted data by shifting one data stream relative to another in time, to remove any astrophysical correlations. The data is taken from GPS time: 815184013-875145614. The search bandwidth considered here is 40-1500 Hz. The analysis is performed using the S5 stochastic analysis pipeline~\cite{PhysRevLett.107.271102}. We parse the time series into 60 second intervals, Hann-windowed, 50\%-overlapping segments, and coarse-grained to achieve a 0.25 Hz resolution. We also mask frequency bins associated with instrumental lines and injected lines for detector calibration and pulsar signal simulation. We apply the stationarity cut described in Ref.~\cite{PhysRevD.76.082003}, which rejects a small percentage (namely, $\sim 3$\%) of the segments. The cross-correlation analysis is performed on each segment of time-shifted data. The outputs from the segments are then combined into a final result and the network maximized-likelihood ratio statistic is computed. We then estimate the GW strain power from the Virgo Cluster.

We assume that the neutron stars in the Virgo Cluster follow the spectral profile VC1 of Fig.~\ref{fig:N_f_virgo}. And the normalization is chosen such that the expected total number of neutron stars is $4\times 10^7$. The distance to the Virgo Cluster is assumed to be $\unit[16.5]{Mpc}$. We estimate the GW energy density using Eq.~\eqref{eq:GWspectrum} and Eq.~\eqref{eq:estimator}.
Following Eq.~\eqref{eq:epsilon} and Eq.~\eqref{eq:epsilon_opt} we estimate one-sigma sensitivity to the average ellipticity of neutron stars in the Virgo Cluster is $\sigma_\epsilon< 9.6\times 10^{-5}$.  

\bibliography{NS_SGWB}

\end{document}